# Assessing the projection correction of Coronal Mass Ejection speeds on Time-of-Arrival prediction performance using the Effective Acceleration Model

**Evangelos Paouris[1], Angelos Vourlidas[1,2], Athanasios Papaioannou[1], Anastasios Anastasiadis[1]**

[1]Institute for Astronomy, Astrophysics, Space Applications & Remote Sensing (IAASARS), National Observatory of Athens, Penteli, GR 15236, Greece.

[2]The Johns Hopkins University Applied Physics Laboratory, Laurel, MD 20723, USA

Corresponding author: Evangelos Paouris (evpaouris@noa.gr)

**ORCID** Numbers:

EP's ORCID: 0000-0002-8387-5202

AV's ORCID: 0000-0002-8164-5948

AP's ORCID: 0000-0003-1682-1412

AA's ORCID: 0000-0002-5162-8821

**Key Points:**

- We introduce a projection correction that results in physically reasonable coronal mass ejection speeds, including halos.

- Corrected CME speeds are 12.8% greater than the plane-of-sky speeds on average.

- Our projection correction in the coronagraph field of view does not significantly improve Time-of-Arrival prediction.

## Abstract

White light images of Coronal Mass Ejections (CMEs) are projections on the plane-of-sky (POS). As a result, CME kinematics are subject to projection effects. The error in the true (*deprojected*) speed of CMEs is one of the main causes of uncertainty to Space Weather forecasts, since all estimates of the CME Time-of-Arrival (ToA) at a certain location within the heliosphere require, as input, the CME speed. We use single viewpoint observations for 1037 flare-CME events between 1996-2017 and propose a new approach for the correction of the CME speed assuming radial propagation from the flare site. Our method is uniquely capable to produce physically reasonable deprojected speeds across the full range of source longitudes. We bound the uncertainty in the deprojected speed estimates via limits in the true angular width of a CME based on multiview-point observations. Our corrections range up to 1.37-2.86 for CMEs originating from the center of the disk. On average, the deprojected speeds are 12.8% greater than their POS speeds. For slow CMEs ($V_{POS} < 400$ km/s) the full ice-cream cone model performs better while for fast and very fast CMEs ($V_{POS} > 700$ km/s) the shallow ice-cream model gives much better results. CMEs with 691-878 km/s POS speeds have a minimum ToA mean absolute error (MAE) of 11.6 hours. This method, is robust, easy to use, and has immediate applicability to Space Weather forecasting applications. Moreover, regarding the speed of CMEs, our work suggests that single viewpoint observations are generally reliable.

## 1. Introduction

Coronal mass ejections (CMEs) are erupting coronal structures observed by white light coronagraphs due to Thomson scattered photospheric light by the free electrons in the corona (Billings, 1966). Since 1996, the Large Angle and Spectroscopic Coronagraph (LASCO; Brueckner et al., 1995) aboard the Solar and Heliospheric Observatory (SOHO; Domingo et al. 1995), has been providing nearly uninterrupted CME observations, recording more than 25000 events to date. Since 2007 we have been able to study CMEs from more than one observational viewpoint revealing essential aspects about their geometry, thanks to the Sun Earth Connection Coronal and Heliospheric Investigation (SECCHI) instruments (Howard et al. 2008) aboard the Solar TErrestrial RElations Observatory (STEREO) mission (Kaiser et al., 2008).

Thompson scattered emission is optically thin and as a result, the features observed in coronagraphs are 2-dimensional projections on the plane-of-sky (POS) of the actual 3-dimensional structures (e.g. Burkepile et al., 2004; Vrsnak et al., 2007). The resulting projection effects lead to underestimation of the speed and overestimation of the CME angular width (Burkepile et al., 2004; Temmer et al., 2009). Projection affects, on one hand, all kinematic and geometric measurements of CMEs limiting our understanding of their physical characteristics and, on the other hand, injects uncertainties in space weather modeling because of uncertainties in the initial speed and half angular width (see e.g. Vourlidas et al., 2019).

Many researchers have tried to correct the geometric and kinematic properties of CMEs to obtain the deprojected direction, speed and angular width using simple geometric assumptions for the CME shape/location (Sheeley et al., 1999; Michalek et al., 2003; Vrsnak et al., 2007; Howard et al., 2008; Thernisien et al., 2006; 2009), proxies, such as shocks (Leblanc et al. 2001,) or multi-viewpoint observations (Howard and Tappin, 2008; Mierla et al., 2008; Temmer et al., 2009; Lee et al., 2015; Jang et al., 2016; Balmaceda et al., 2018).

It is reasonable to expect that multi-viewpoint observations will result in better speed estimates, as long as the CME 3D shape can be reliably reconstructed, which is not always the case (e.g. event overlap, small angular separation between viewpoints, etc.). This is basically the take away message from the review of Time-of-Arrival (ToA) measurements over the last years in Vourlidas et al. (2019). A bigger concern, from a space weather perspective, is the lack of guarantee that such measurements will be available in the future. It is uncertain, therefore, whether multi-viewpoint methods will continue to benefit space weather research in the long run. On the other hand, single-viewpoint observations, from the Sun-Earth line, are poised to continue with the scheduled deployment of operational coronagraphs in the (Geostationary Operational Environmental Satellite-S) GOES-S and (Space Weather Follow-On) SWFO spacecraft starting in 2024. Hence, it is worthwhile to investigate methods to improve the reliability of CME speeds extracted from single-viewpoint measurements and to assess the impact of such improved speeds on ToA performance. Any improvement in the CME ToA forecasting accuracy should benefit both CME research and operations in the near future.

We are not seeking to evaluate the accuracy of CME speeds from a single viewpoint. Such works have been published already (see Temmer et al., 2009; Gao and Li, 2010; Bronarska & Michalek 2018, and references mentioned before). Our goal is to evaluate the effect of a particular speed correction as input to ToA empirical models in preparation for a future scenario with CME observations from along the sun-earth line only (e.g. from SWFO or GOES-S). For this reason, we use the standard empirical approach for deprojecting the speed (the ice-cream model, or more specifically, a shallow ice-cream model). We also take advantage of STEREO observations, albeit indirectly, by means of the width estimate.

We use an existing database of CME measurements (Section 2) associated with flares (for the CME source region) to reevaluate the assumptions that go into the geometric correction methods (Section 3). We then proceed with an improved geometric correction (section 4) that avoids speed overcorrections, especially for halo-CMEs, that have plagued previous similar efforts. We examine the performance of this method on CME ToA in Section 3.4. We conclude in Section 5.

## 2. Data Sources and Methodology

The CME information utilized in this work is obtained from the SOHO/LASCO coordinated data analysis workshop (CDAW) database ([https://cdaw.gsfc.nasa.gov/CME_list/](https://cdaw.gsfc.nasa.gov/CME_list/)) that covers the period between January 1996 and June 2017. Soft X-ray data from Geostationary Operational Environmental Satellites (GOES) are obtained from National Geophysical Data Center (NGDC) ([ftp://ftp.ngdc.noaa.gov/STP/space-weather/solar-data/solar-features/solar-flares/x-rays/goes/xrs/](ftp://ftp.ngdc.noaa.gov/STP/space-weather/solar-data/solar-features/solar-flares/x-rays/goes/xrs/)). In particular, 22939 solar flares and 28718 CMEs were identified in the January 1996 – June 2017 period.

Sheeley et al. (1999) used an innovative method for the calculation of the speed and acceleration of CMEs using a slice of a coronagraph image along a predefined radial path and with several time-lapse coronagraph images created time-distance plots of propagating disturbances. The heliocentric distance of a feature (e.g., CME leading edge or core) as a function of time, combined with simple quadratic fitting in most cases, led to the calculation of the speed and acceleration. Considering how the expansion and outward motion of a gradually accelerating CME is contributing to the projected, on the plane of sky, speed they assumed that the CME consisted of a thin spherical shell. According to this assumption, the expansion and radial velocity of a bubble with its center in the Sun-Earth line are associated with the relation:

$$V_{rad} = \frac{V_{EXP}}{\sin a}$$

*Eq. 1*

where *a* is the angular half width defined as the angle between the radial from the Sun center through the center of the expanding bubble and the tangent line through the Sun center. This relation indicates that, for a typical value of $a = 20°$, the radial speed is almost three times the expansion speed relative to the bubble center.

Sheeley et al., (1999) concluded that geometric effects alone could not account for the observed acceleration and deceleration of halo CMEs.

Leblanc et al. (2001) assumed that CMEs erupt radially from the flare site, so they used the coordinates of the parent solar flare to deproject the POS speed and obtained the radial speed from **Eq. 2**

$$V_{rad} = V_{POS} \frac{1 + \sin a}{\sin \varphi + \sin a}$$

*Eq. 2*

where $a$ is the angular half width of the CME (as defined above) and $\varphi$ is given by $\cos\varphi = \cos\theta \cos\lambda$,

where $\theta$ is the longitude, and $\lambda$ is the latitude of the associated solar flare. A more intuitive way to discuss projection effect is the plane-of-sky (POS) angle $\psi$,

$$\sin \psi = \sin \left( 90° - \varphi \right) = \cos \varphi$$

*Eq. 3*

Leblanc et al. (2001) report that for an angular half width $a = 30°$ and $\varphi = 0°$, $30°$, $60°$ and $90°$ the ratio, $V_{rad}/V_{POS}$ becomes 3.0, 1.5, 1.1 and 1.0, respectively. $\varphi = 0°$ indicates a CME originating from the center of the visible solar disk, while $\varphi = 90°$ represents a limb event. This method is straightforward but suffers from overcorrection for events near the center of the visible solar disk. For example, for a fast CME near the disk center ($\varphi \approx 0°$) with an angular half width of $a = 30°$ and a $V_{POS}$ of 1500 km/s, the $V_{rad}$ obtained from this method is $\approx 4500$km/s - which is unrealistic.

Michalek et al. (2003) applied a method for correcting the angular width, velocity, and source location of 72 Halo CMEs during the period August 1996 – December 2000. They found that the radial speed ranged from 95 km/s up to 2590 km/s, and the ratio of $V_{rad}/V_{POS}$ ranged from 0.59 up to 2.97. In 21 cases, the ratio was less than 1.0, indicating that the radial speed was actually less than the POS speed measured in the LASCO coronagraphs, which is difficult to reconcile with the expectation from a POS projection (i.e. the projected speed is always less than the radial one). In any case, the linear fit of the radial speeds as a function of POS speeds, using the data from columns 3 and 11 from Table 1 of Michalek et al., (2003), is

$$V_{rad} = 1.022 \cdot V_{POS} + 142.8 \text{ [km/s]}$$

*Eq. 4*

This relation (**Eq. 4**) is very similar to the relation of Paouris and Mavromichalaki (2017b) obtained from a sample of 87 CMEs with angular widths w < 360° (no full halo events) associated with solar flares. Their equation reads

$$V_{rad} = 1.027 \cdot V_{POS} + 41.5 \text{ [km/s]}$$

*Eq. 5*

Michalek et al. (2003) conclude that the projection effects are important and find that the corrected speeds are on average 20% larger than the POS velocities.

Another technique for the correction of the projection effects was presented by Howard et al. (2008). This method was applied automatically in a sample of 10000 CMEs covering the period 1996-2005 from the CDAW catalog. From the full sample, they associated 1961 CMEs with a source event, i.e., solar flare. They used the elongation ε, which is the angular distance between Sun center and the outmost point (P point) of the transient in distance-time plots (for the P point approximation see Howard et al. 2006), and considering the CME trajectories in a 3-D geometry established a correction for the projection effect. They found that $V_{rad}/V_{POS}$ ranged between 1.7 and 4.4. They also found many events with corrected speeds > 3000 km/s, or even > 5000 km/s, which seem unphysical.

Speed overcorrection is of serious concern for such correction schemes. From a total of 29714 CMEs recorded in the CDAW database between January 1996 – March 2018, only 2 CMEs have projected speeds above 3000 km/s, i.e. events on 10/11/2004 ($V_{POS}$ = 3387 km/s) and 10/09/2017 ($V_{POS}$ = 3163 km/s). Assuming CMEs are uniformly distributed in longitude, there could be as many as 10000 CMEs within ±30° longitude from the limb that hence exhibit minimal projection effects. If there were CMEs faster than 3400 km/s, they should have been found. This is an important issue as it pertains to upper limits of CME energetics. The largest CME mass in the CDAW database (Vourlidas et al., 2000) is $2*10^{17}$ gr. Assuming an upper limit for the CME speed of ~3500 km/s, the largest possible CME kinetic energy is $1.23*10^{34}$ ergs, which is close to the upper estimates of the available free energy in active regions (Emslie et al. 2012). Higher kinetic energies would be energetically impossible. Therefore, methods that result in speeds in excess of 4000 km/s are likely unreliable.

In a recent work, Balmaceda et al. (2018), analyzed a sample of 460 events via the use of a supervised image segmentation algorithm (see also Vourlidas et al., 2017) which were measured in observations from the two COR2 telescopes onboard the STEREO mission (STEREO A and B) from May 2007 up to September 2014. They found that projection effects are more important for the determination of CME's angular width rather than for CME's speed. The projected speeds were, on average, underestimated by at least 20% when compared with the deprojected speeds obtained from a triangulation method using observations from both STEREO A and B satellites.

### 3. Analysis

### *3.1 Association of a CME with a solar flare*

Our goal is to develop a geometrical method for the correction of the projected CME speed. We start by obtaining positional information for 22,939 solar flares ($\geq$C1.0), distributed as 20630 C-class, 2138 M-class and 171 X-class, and 28,718 CMEs covering the period January 1996 – June 2017.

First, we find the latitude and longitude of the source region of the CME. To achieve this, we associate a CME with a solar flare and use the flare's latitude ($\lambda$) and longitude ($\theta$). Due to the large number of events, it is impractical to perform the association of a CME to a solar flare by visual examination of the CME and soft X-ray series. Instead, we apply temporal and spatial criteria similar to those used by other researchers (see e.g. Vrsnak et al., 2005; Youssef, 2012; Papaioannou et al., 2016; 2018a; 2018b, and references therein). From the CDAW database, we obtain the time ($t_1$) when the CME first appears in LASCO C2 coronagraph, and estimate the lift-off time ($t_0$) using a simple back-extrapolation method of its trajectory to the Sun's surface:

$$t_0 = t_1 - \frac{\Delta x}{\upsilon}$$

*Eq. 6*

assuming that the CME speed, $\upsilon$, is constant from the surface to the distance of the first measurement of LASCO C2 field of view (FOV) with $\Delta x$ usually of the order of $3R_{Sun}$ (Vrsnak et al., 2005). Once $t_0$ is estimated, we associate a solar flare to that CME, under the condition that it occurs within $\pm 1.5$ hours (*temporal criterion*). In some cases, there were multiple flares inside the $\pm 1.5$ hours window, especially around solar maximum. To overcome this, we apply a *spatial* *criterion*. For each solar flare, we calculate the position angle PA$_{SF}$ as this is inferred by its heliographic coordinates using the equation:

$$PA_{SF} = \tan^{-1}\left(\frac{\sin\theta}{\tan\lambda}\right)$$

*Eq. 7*

where θ and λ are the flare heliographic longitude and latitude, respectively, and compare it to the CME angular width (*w*) and position angle PA$_{CME}$ from the CDAW database. Our spatial criterion requires that the position angle of the flare should be within the position angle interval spanned by the CME (see Vrsnak et al., 2005; Youssef, 2012).

The combination of both *temporal* and *spatial* criteria leads to a final sample of 1037 CME-solar flare pairs, including 69 X-, 275 M- and 693 C-class solar flares. This is the sample we analyze in the rest of the paper. We recognize it is clearly biased towards energetic events, originating from active region complexes. Those are, however, the most space weather-relevant events, which are our focus.

### 3.2 A new geometrical approach

As we discussed earlier, various attempts to correct for projection effects on CME speeds have taken place, but they all suffer from overcorrection. The same problem affects the correction approaches relying on the assumption of radial propagation from the flare site. To find a solution, we come back to the fundamental question of speed measurements: *what is the nature of the front measured in the coronographic images*?

The convention is to measure the outmost front in a given image, and follow it (along the same position angle, to the maximum possible extent) as the CME moves across the FOV of the coronagraph. The underlying assumption is that the outermost front corresponds to the nose of the CME (otherwise, the radial propagation from the flare site is no longer a valid assumption) or at least to the nose of the shock driven by the CME. But, *how valid is this assumption and under what conditions?*

Leblanc et al. (2001) took this into account to derive **Eq. 2** under the assumption that the CME has the shape of an 'ice-cream' cone. The ice-cream cone structure is a widely-used assumption in CME research (e.g. Leblanc et al., 2001, Zhao et al., 2002; Michalek et al., 2003; Xie et al., 2004;

Xue et al., 2005; Makela et al., 2016 and references therein) because it seems to capture rather well the spherical shape of the 3-part CME. However, it leads to rather unrealistically high speed corrections for events close to the central meridian, which led us to question whether the assumption of a spherical front is valid for all CMEs. Recently, Balmaceda et al (2018) reported that CMEs with a characteristic loop front, called 'L-CMEs' have much higher expansion speed than flux-rope type events (called 'F-CMEs', which tend to correspond to the classic 3-part events). L-CMEs are more common in halo events. Based on their appearance and their lateral expansion, such events significantly deviate from the spherical, 'ice-cream' cone morphology and should be better described by a shallow-cone model (a frustrum of a spherical cone, to be precise). Hence, we investigate the effect of this assumption on speed deprojection.

Let us consider the CME as an expanding shallow cone with angular half width, $a$, propagating at an angle $\psi$ from POS (**Figure 1**). The CME is associated with a solar flare and expands radially from the flare location with known heliographic latitude, $\lambda$, and longitude, $\theta$. The figure suggests that, for certain propagation directions, the CME outmost point as seen in coronagraphs is actually the CME flank and not the CME nose.

If the CME nose is located at heliocentric distance $r_1$, and the CME flank at $r_2$, then the projected nose and flank heights, $r_n$ and $r_{fl}$, will be

$$OA = r_n = r_1 \cdot \cos\left(\psi\right)$$

*Eq. 8*

and

$$OB = r_{fl} = r_2 \cdot \cos\left(\psi - \alpha\right)$$

*Eq. 9*

For the shallow cone model the CME front (red arc in **Figure 1**) is an arc of angular width w. Therefore, the distances $r_1$ and $r_2$ are both radii of the same circle, $r_1 = r_2 = R$

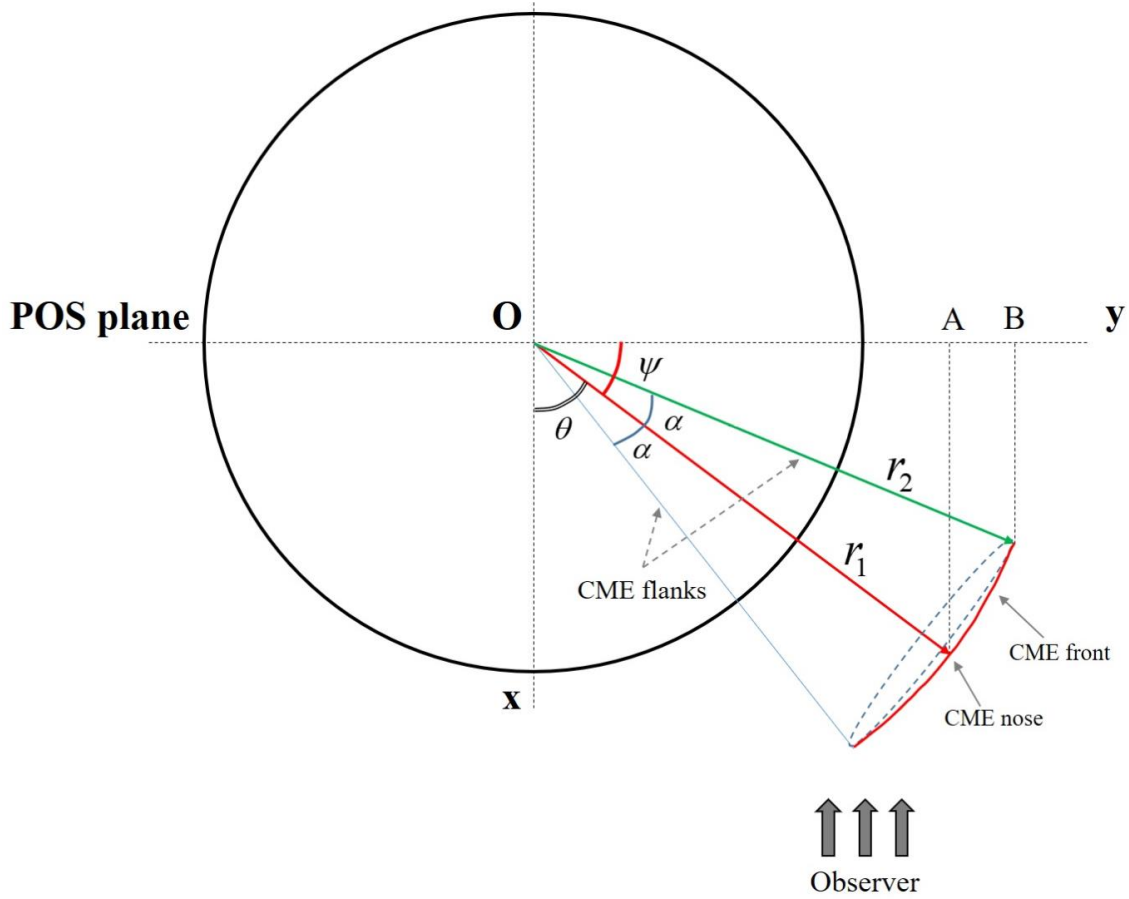

*Figure 1: A CME of angular half width a is propagating at an angle ψ from POS, radially from flare location of LAT = 0° and LON = θ°. The observer is on the xy plane. The POS plane is normal to the y axis. Observing the CME from above (North Pole of the Sun).*

The ratio of $r_{fl}/r_n$ from **Eq. 8** and **Eq. 9** is:

$$\frac{r_{fl}}{r_n} = \cos(\alpha) + \tan\psi \cdot \sin(\alpha)$$

*Eq. 10*

This ratio should be < 1, for the assumption that the outmost projected point of a CME is the projection of its nose, to be valid. However, for an average CME width of 50°, $r_{fl}/r_n > 1$ for ψ >

12.5°. Therefore, even for very small departures from POS propagation, the outmost CME point on an image *may not* be the CME nose. For wide CMEs, the situation improves, but only marginally. For example, for a 90°-wide CME $r_{fl}/r_n > 1$ for $\psi > 22.5°$. In our opinion, this width value is a likely upper limit for the true width (see Balmaceda et al. 2018) as coronagraph observations can be influenced by misidentification of the shock, streamer deflections, etc. and this can lead to an incorrect estimation of the CME width. Even for a 120°-wide CME, $r_{fl}/r_n > 1$ for $\psi > 30°$. Therefore, we conclude that the outermost point measured in CMEs tends to correspond to the CME flank rather than the nose, for CMEs consistent with the shallow cone description.

The case $r_{fl}/r_n = 1$ holds for $\psi = \dfrac{\alpha}{2}$ which is actually the one-quarter of the angular width w of the CME. So, our assumption $r_{fl}/r_n < 1$ holds when $\psi < \dfrac{\alpha}{2}$ or using the known coordinates of the associated solar flare with latitude λ and longitude θ we get:

$$\psi < \frac{\alpha}{2} \Leftrightarrow \sin\psi < \sin\left(\frac{\alpha}{2}\right) \Leftrightarrow \cos\varphi < \sin\left(\frac{\alpha}{2}\right) \Leftrightarrow$$

$$\cos\theta \cdot \cos\lambda < \sin\left(\frac{\alpha}{2}\right)$$

*Eq. 11*

This equation defines the angular limit for the flank-to-nose correction. For example, if the true angular width of a CME is 50° then $\psi = 12.5°$. Therefore, no flank correction is needed if the CME is associated with a solar flare of POS angular less than 12.5° because in this case we have $r_{fl} < r_n$. Evidently the (*true*) angular width of the CME is a key parameter for deciding whether a flank correction is required.

The observed angular width, $w_p$, is a projection of the true angular width w and is generally difficult to estimate without two or three viewpoint reconstructions. This approach is also impractical for space weather applications. Instead of trying to de-project the width on an event-by-event case, we bound its possible values based on previous multi-viewpoint work. First, we consider that the large apparent CME widths (say, above 90° or so) are results of projections and/or structure overlap (e.g.

shock or deflected streamer signatures rather than true width. Shen et al., (2013) studied 86 full Halo CMEs from the CDAW database and applied the Graduated Cylindrical Shell (GCS) model (Thernisien et al., 2006) to each one. They concluded that the deprojected angular width varies within a large range (44° - 193°) implying that the projection effect is a major reason leading a CME to appear as halo. However, we note that the upper estimates of the deprojected widths are inconsistent with the relatively small extent of CME-associated dimmings (of the order of an AR size, see Aschwanden et al 2009). In reality, many of the observed large widths are the results of the shocks ahead of the CME rather than the CME itself, in most cases (Vourlidas et al. 2013, Kwon et al. 2015). For example, there are many cases of halo CMEs detections from L1 (LASCO) where the same CME has a much smaller width from a different perspective, e.g. STEREO-A or -B (Vourlidas et al. 2017; Balmaceda et al. 2018).

These works introduced the Multi-Viewpoint CME (MVC) catalog which is the only database of CMEs to date to uniquely identify events from two different observational viewpoints (STEREO A and B) and to include a morphology classification for each CME. The classification aims to identify events consistent with white light signatures expected for a magnetic flux rope (e.g. circular or cylindrical shapes, evidence of striations within the CME body and so on; please see Vourlidas et al. (2017) for details). These CMEs are classified as L- and F-types in Vourlidas et al. (2017). L-type CMEs are interpreted as F-type CMEs seen face-on. We consider only these two types as truly representative CMEs, since they are consistent with expectations from all theories of CME initiation (e.g. Vourlidas et al. 2013; Patsourakos et al. 2020). Since they represent the two extreme projections of a magnetic flux rope (F-CMEs have their current axis into the sky plane; L-CMEs have their current axis along the sky plane), their measured widths should capture the range of true CME widths, at least under the above assumption. So we proceed as follows.

We take the mean values of the angular width of L- and F-CMEs from Table 2 of Balmaceda et al., (2018). To remain consistent with the assumption that L-CMEs are CMEs seen face-on, we take the maximum angular width (mean width plus one standard deviation) of all L-Type CMEs as upper limit for our work ($w_L = w_{upp} = 94°$). Hence the F-CME minimum angular width (mean width minus one standard deviation) is the lower limit ($w_F = w_{low} = 41°$). In other words, we assume that the true widths of all CMEs range from 41° to 94°. Now, we can determine the source longitudes beyond which the flank correction becomes insignificant. According to Eq. 11 they range from $\psi = 0°$ (limb events) to $\psi = 10.25° \approx 10°$ (for F-type CMES, $w_F = 41°$) or to $\psi = 23.5° \approx 24°$ (for L-CMEs, $w_L = 94°$).

We can now approach the deprojection of $V_{POS}$ from a different perspective. We assume a CME radially expanding along OM (***Figure 2***), as a shallow cone of angular width, $w_{lim.}$ The CME is further associated with a flare at POS distance, $\psi$. The y-z plane is the POS of the coronagraph. The projection of the radial OM on the POS correspond to the usual projected height-time (h-t) measurements used for kinematics calculations. This projected point is the outmost point of the CME (the so-called 'front') seen in the coronagraph images. As we argued earlier, this point will tend to correspond to the CME flank nearest to the POS (except when the CME propagates along the POS). The angular location of that point is

$$\psi_{cor} = \psi - \frac{w_{lim}}{2}$$

*Eq. 12*

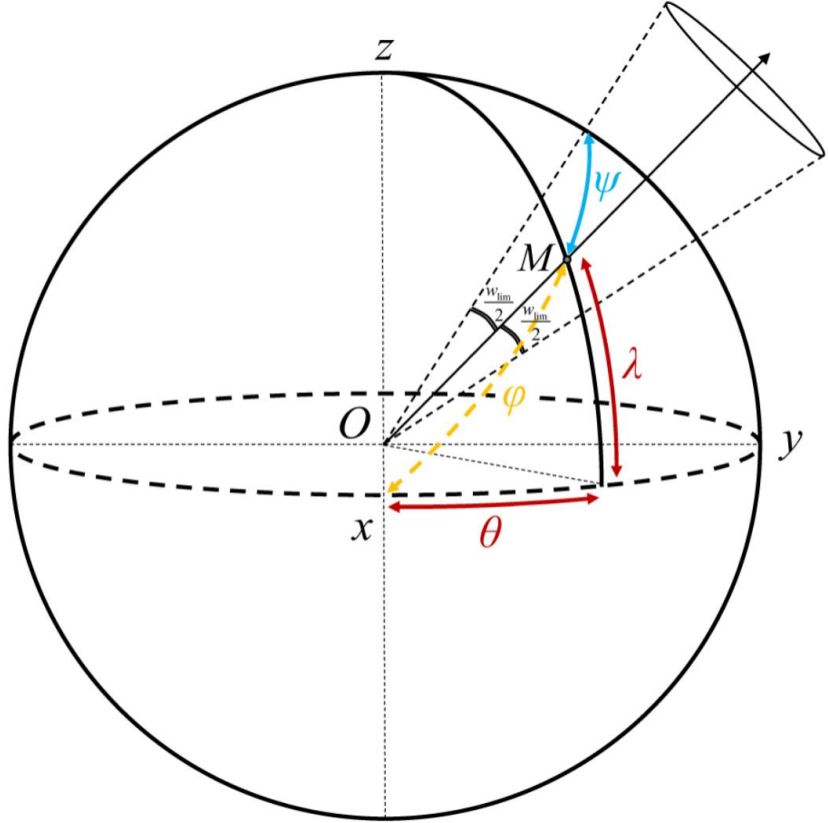

**Figure 2:** *Geometrical diagram of an expanding CME radially from the associated solar flare location (point M) of longitude θ and latitude λ. The observer is on the Ox line. The line OM is the cone axis of the expanding CME of angular width w$_{lim}$. The POS angle ψ and the complementary angle φ are also indicated.*

The radial distance of OM (*r*) can then be derived from the equation:

$$r = \frac{h}{\cos \psi_{cor}}$$

*Eq. 13*

with ON (*h*) being the height values from the h-t measurements of each CME and ψ$_{cor}$ is the corrected POS angle.

The correction of **Eq. 13** applies only for source locations where the CME "front" corresponds to the CME flank i.e. where **Eq. 11** stands. For the other locations we use the POS angle $\psi$ without the "flank correction":

$$r = \frac{h}{\cos \psi}$$

*Eq. 14*

The range of applicability is given by **Eq. 11** for the two width limits, is $\pm 10.25°$ for the lower angular width limit and $\pm 23.5°$ for the upper limit.

To summarize our correction scheme:

- Instead of using the projected CME widths, we assume that the true widths vary between $41°$ and $94°$, which are the average widths minus/plus one standard deviation of F- and L-CMEs according to Balmaceda et al., (2018).

- We deproject the h-t measurements under the assumption of a shallow cone (a cone frustum) using **Eq. 13** for CMEs with POS angle $|\psi| < 10.25°$ ($23.5°$) for upper (lower) width limit and use **Eq. 14** for the remaining cases.

- We apply this scheme to all h-t measurements for the 1037 CMEs in our sample. Then, we use a linear and a quadratic fit to obtain the radial speed, $V_{rad}$, and the radial speed at 20 $R_{Sun}$ ($V_{20R}$).

The resulting ratio $V_{rad}/V_{POS}$ ranges from 1.0 to 2.86, for the lower limit of the angular width, and from 1.0 to 1.37 for the upper limit. The events are presented in ***Figure 3*** as a function of their position on the solar disk, using the same color code for upper and lower limits of ratio $V_{rad}/V_{POS}$. It is evident that projection effects are important close to the center of the solar disk while on sites with absolute POS angle $|\psi| < 24°$ (***Figure 3*** (a)) and $|\psi| < 10°$ (***Figure 3*** (b)) projection effects are insignificant. Another important result concerns the radial (deprojected) speeds: the corrected speeds in this work range from just a few km/s ($\approx$40 km/s) up to 3960 km/s providing meaningful limitations to the deprojected CME radial speeds, in contrast with previous works (see Section 2). In particular, for non-halo CMEs (with projected angular width < 180°) the deprojected speeds range from $\approx$40 km/s up to $\approx$2270 km/s. For halo CMEs ($180° \leq w_p \leq 360°$) the LASCO linear speed ranges from ≈130 km/s up to ≈2660 km/s, while the largest deprojected speed, for the lower width limit of 41°, range from ≈200 km/s to ≈3960 km/s. The largest corrected speed of 3961.0 km/s, using the lower limit of width, corresponds to the full halo CME of October 28[th] 2003, associated with an X17.2 class solar flare at heliographic coordinates S16E08, very close to the center of the solar disk. According to the CDAW database, this CME had a linear speed of 2459 km/s and as result the ratio $V_{rad}/V_{POS}$ is ≈ 1.61. The corrected speed using the upper width limit for the same event is 2714.7 km/s. The minimum and maximum corrected speeds (using the upper and lower width limits) are the range for each point (error bars) in **Figure 4**.

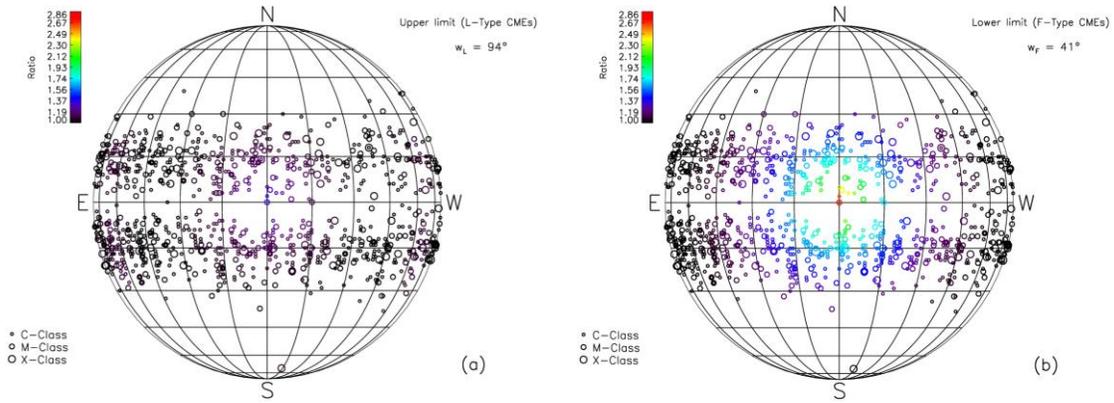

*Figure 3: Ratio of the deprojected speed to the POS speed for (a) the upper limit (w$_L$=94°) and (b) the lower limit (w$_F$=41°) of angular width for 1037 CMEs associated with solar flares.*

The lower and upper limits of the angular width used to define the range of the corrected speed in each case, presented in **Figure 4** as a function of the POS speed for our sample of 1037 events. The solid red line is the derived empirical relation between the deprojected (radial) and POS speeds:

$$V_{rad} = 1.128 \cdot V_{POS} \ [km/s]$$

*Eq. 15*

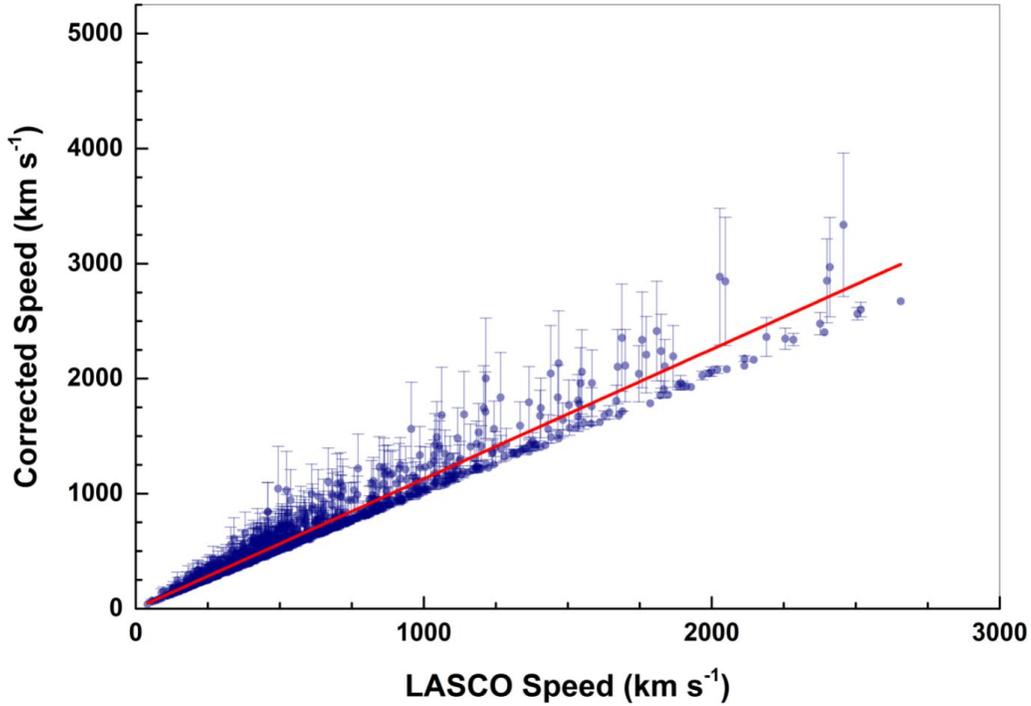

*Figure 4: Deprojected against POS speeds plot for the 1037 flare-associated CMEs.*

The Pearson correlation coefficient between the corrected and the POS speeds is 0.99 and the slope of the red line (1.128±0.005) implies that the radial speeds are on average ≈12.8% greater than the POS speeds. It is evident from **Figure 4**, that our method provides reasonable corrected speeds. None of our corrections (mean values) exceeds ~3400 km/s, which is the fastest (projected) speed recorded directly by the LASCO coronagraphs. This is a very encouraging result, providing confidence in approaching the speed correction problem with the shallow cone model. We compare our technique to other approaches in Sec. 3.3 but first we examine in some more detail the degree of correction for different speeds.

We split the sample in three bins based on the LASCO linear speed of slow ($V_{POS}$ < 500 km/s), fast (500 km/s < $V_{POS}$ < 1500 km/s) and very fast ($V_{POS}$ > 1500 km/s) CMEs and we calculate the ratio between the POS speed and the minimum/maximum corrected speeds. One of the advantages of our study is that we apply our methodology directly on the h-t measurements for each CME. This allow us to assess much information, such as the number of h-t measurements (NHT) for each event

- which we will use to evaluate and fine-tune our technique. At this point we clarify that the geometric corrections could be equally applied to POS speed directly. The main reason for using our method directly to the h-t measurements is to experiment with other approaches and study e.g. the effects from using speeds at various distances from 15 to 30 $R_{Sun}$. Another reason for the use of h-t points is to evaluate the reliability of speeds as a function of the available number of h-t points as we discuss in the next paragraph.

For example, we compare the correction ratio between the POS and corrected speeds for the whole sample of 1037 events against the sample of events (646) with $\geq$ 10 h-t points, which may have more reliable speed fits. The number of events per bin and the correction ratios as a percentage difference are shown in **Table 1**. We derive two findings from this table. The first, and most important one, is that the mean correction is very similar for both samples and hence the h-t fits are generally reliable, irrespective of the number of h-t points. The second finding is a tendency for the correction ratio to narrow with increasing CME speed. This is a subtle effect and may not be important. It suggests, at least to us, that faster CMEs are likely wider, with more uniform speeds across their fronts than narrower CMEs. The speed uniformity is consistent with the fronts being shocks rather than CME fronts, for these events.

*Table 1*. *Mean correction ratios for slow, fast and very fast CMEs.*

|  | Number of events | Speed bins [ , ) in km/s | Mean correction (%) |
|---|---|---|---|
| All events | 536 | [0, 500) | 5 - 27 |
|  | 443 | [500, 1500) | 5 - 24 |
|  | 58 | > 1500 | 4 - 19 |
|  | Total: 1037 |  |  |
| Events with HT ≥ 10 | 361 | [0, 500) | 5 - 28 |
|  | 278 | [500, 1500) | 4 - 22 |
|  | 7 | > 1500 | 4 - 14 |
|  | Total: 646 |  |  |

### 3.3 Comparison with other correction methods

A summary of our results from the new geometrical method (correcting the height/time measurements) and the results from similar methods from other researchers is presented in **Table 2**.

*Table 2. The ratio of the radial to POS speed as a metric for various works on the field of correction of the linear speed from projection effects.*

| Research | Comment | Ratio $V_{rad}/V_{POS}$ |
|---|---|---|
| Sheeley et al., 1999 | Geometrical method based on the half-angular width ($\alpha$) | r≈3 (for $\alpha$ = 20°) |
| Leblanc et al., 2001 | Geometrical method based on the half-angular width ($\alpha$) and the angle of active region from disk center ($\varphi$) $\varphi$ = 0° when CME originates from disk center and $\varphi$ = 90° for limb events | r≈3 (for $\alpha$ = 30° and $\varphi$ = 0°) |
| Michalek et al., 2003 | Geometrical method with CME cone model mean error 20% | 0.6 < r < 3 |
| Howard et al., 2008 | Geometrical method using elongation, width, $\theta$ and $\lambda$ | 1.7 < r < 4.4 |
| Balmaceda et al., 2018 | Mean error 20% | - |
| Our method | Geometrical method using half-angular width limitations, $\theta$ and $\lambda$, h-t measurements Upper limit for angular width $w_L$ = 94° Lower limit for angular width $w_F$ = 41° Mean error 12.8% | 1.0 < r < 1.37 1.0 < r < 2.86 |

At this point, we focus on the comparison between our method and Leblanc's method because the two methods are similar, differing only in the CME cone model used. Leblanc et al. (2001) assume that the CME front is described by a full ice-cream cone model while our method assumes a shallow cone model. Leblanc's method results in unreasonable speeds especially when using the projected half-angular width of CME. We applied our assumptions for the true angular width between 41° and 94° to the Leblanc et al. method. The results are presented in **Figure 5**.

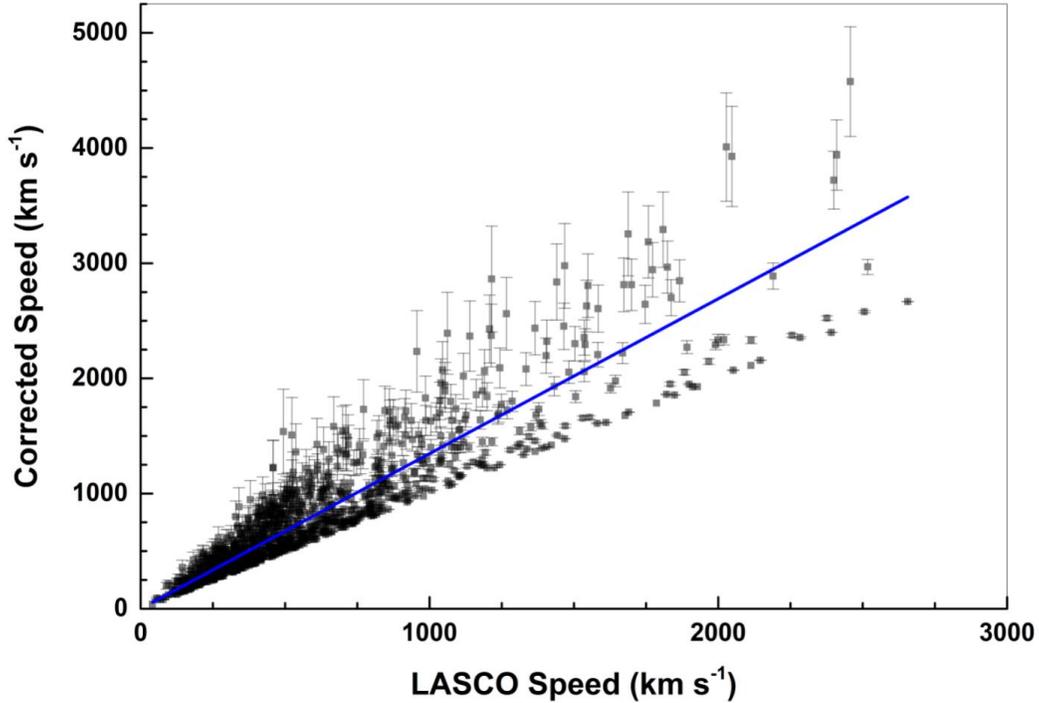

*Figure 5: Deprojected against POS speeds plot using Leblanc et al (2001) method in our sample.*

The solid blue line is the derived empirical relation between the deprojected (radial) and POS speeds using Leblanc's method:

$$V_{r(Leb)} = 1.346 \cdot V_{POS} \ [\text{km/s}]$$

*Eq. 16*

The Pearson correlation coefficient between the corrected and the POS speeds is 0.97 and the slope of the line (1.346±0.010) implies that the radial speeds are on average ≈34.6% greater than the POS speeds. Hence, this method results is almost 3 times greater speed corrections than our method. In particular, the maximum CME speeds using Leblanc's method are 5053 - 4100 km/s. The impact of each method in the calculation of the Time-of-Arrival (ToA) of a CME to Earth is discussed in the next session.

### 3.4 Application to the Time-of-Arrival problem

Two major questions arise when a CME is detected in coronagraph images: a) whether the CME will hit the Earth or not and b) if so, when the impact will take place, i.e. what is the time of arrival (ToA). We use our corrected speed to focus on the second question. We take a sample of 192 CMEs/Interplanetary CMEs with known flare source region (Paouris and Mavromichalaki, 2017a) that have been used as seeders to the Effective Acceleration Model (EAMv2) (Paouris and Mavromichalaki, 2017b). We obtain the LASCO speeds and correct them using our method to derive an upper/lower limit speed. The speeds and the time of first appearance of each CME in LASCO coronagraphs are then used independently as input to EAM to derive the predicted ToA from the model. The actual arrival time is obtained from the Advanced Composition Explorer (ACE) in-situ measurements at the time of the shock. From that, we derive the error, which is the difference between the predicted and the actual arrival time:

$$\Delta t_{err} = t_{predicted} - t_{observed}$$

*Eq. 17*

We use statistical metrics to compare the ToA performance for the three speed inputs. These metrics are: a) the mean absolute error (MAE),

$$MAE = \frac{\sum |\Delta t_{err}|}{n}$$

*Eq. 18*

b) the mean error (ME)

$$ME = \frac{\Delta t_{err}}{n}$$

*Eq. 19*

and c) the root mean square error (RMSE)

$$RMSE = \sqrt{\frac{\sum \left( \Delta t_{err}^2 \right)}{n}}$$

*Eq. 20*

where n is the sample size of 192 events.

The obtained metrics are presented in **Table 3**. We compare the performance of the Leblanc et al (2001) method to ours using both their assumption (all CMEs have the same 36° angular half width) and our assumption (CME width lies between 41° and 94°). The standard errors for LASCO and Leblanc et al (2001) 'default' methods are calculated by a simple bootstrap method with replacement for $10^6$ runs. The standard errors for the methods using the angular width limits (upper/lower) are derived directly from the range of each metric.

*Table 3. Statistical ToA metrics (in hours) for POS and corrected CME speeds derived by applying the Leblanc et al and our methods on a sample of 192 events adopted from Paouris and Mavromichalaki, (2017b).*

|  | **LASCO** | **Leblanc et al. (2001) method** | | **Our method** |
|---|---|---|---|---|
|  | **POS speed** | **Default (w = 36°)** | **Using angular width limits** | **Using angular width limits** |
| MAE | 17.80 ± 0.87 | 19.41 ± 0.93 | 19.99 ± 1.19 | 17.71 ± 0.37 |
| ME | 2.04 ± 1.54 | -12.15 ± 1.43 | -13.03 ± 2.30 | -4.22 ± 4.32 |
| RMSE | 21.52 ± 0.92 | 23.31 ± 1.03 | 23.88 ± 1.23 | 21.55 ± 0.44 |

Overall, there is a significant improvement in the ToA metrics using our speed compared to the Leblanc et al. (2001) method and a slight improvement on MAE compared to the LASCO POS speeds. In particular, we obtain lower MAE and narrower errors for MAE and RMSE. The uncertainty ranges overlap for corrected and uncorrected speeds so the improvement is marginal, at least for the set of the 192 CMEs considered. However, the results are encouraging and remains to be seen if the improvement will hold when applied to larger or different samples. We discuss such an attempt later in this section.

Our method tends to result in slightly earlier CME arrivals compared to slightly later arrival with POS speeds. What is clear from **Table 3** is the inferior performance, across all metrics, of the circular ice-cream model assumption in the Leblanc et al (2001) method. The method results in statistically higher MAEs and significantly earlier CME arrivals. This is the result of the overcorrection in the speeds, which we discussed before. Similarly, the slight improvement from our method is due to the rather modest speed corrections under the shallow cone model assumption, which are ≈ 12.8% on average.

We note that the analysis is based on the empirical EAMv2 model. It assumes, as do other similar empirical models, that: (1) a smooth deceleration curve (a polynomial in the EAMv2 case) describes the interplanetary propagation of the CME, and (2) that the speed derived in the coronagraph FOV is the speed component along the Sun-Earth line. We explored, to some extent, the latter assumption here. We examined the effects of the CME front shape and direction in an empirical deprojection of the speeds. Clearly, as the ToA metrics in **Table 3** suggest, there is little benefit to be had, even though we strove to apply a consistent set of assumptions based on the most recent understanding of CME shape and 3D widths. One could question the assumption of radial propagation from the flare site. It is not particularly strong, since non-radial propagation has been observed even for active region CMEs (e.g. see Liewer et al. 2015, and references therein). The non-radiality is hard to assess, however, for events propagating towards the observer from observations along the Sun-observer line (the focus of this paper). Also, the more energetic eruptions (resulting in higher speeds) are more likely to have higher energy densities than the surrounding corona and hence are less likely to deflect. In any case, our shallow cone assumption reduces the effects of radial deviations because it implies a flatter speed profile across the CME front. This is a reason why we obtain reasonable deprojected speed with this assumption. Therefore, we conclude that, although the non-radiality of CMEs affect the accuracy of the correction, it is unlikely to be the dominant effect for the large MAEs.

It is far more likely that the uncertainty arises from the interplanetary (IP) CME propagation, which is not well described by the smooth deceleration assumption. Although far from established (Vourlidas et al 2019), there are several indications that CMEs undergo various kinematic changes during their IP evolution that may induce both acceleration and deceleration or even nothing at all (e.g. Colaninno et al 2013; Wood et al. 2017). Therefore, we conclude that efforts to derive more nuanced empirical propagation models should improve the MAE performance of single-viewpoint ToA predictions much more than efforts to deproject CME speeds measured in coronagraphs (a similar conclusion on the utility of coronagraph measurements was reached by Colaninno et al. 2013). This is, however, outside the focus of this paper.

So, what can we do with the existing models and approach? We mentioned above that fast CMEs are more likely to expand radially from the flare site. **Table 1** indicates that faster CMEs may require less correction for projection effects. On the lower speed end, Colaninno et al (2013) and Vourlidas et al. (2013) found that the kinematics of events with speeds close to the ambient solar wind speed are more uncertain because of the higher degree of interaction with the ambient structures. These findings indicate that the speed of the event in the coronagraph FOV may have an effect on the accuracy of the empirical ToA determinations; e.g. faster events should perform different than slower ones. To investigate this, we split our sample of 192 CMEs into eight speed bins containing 24 events each and examine the MAE performance as a function of the POS speed (**Figure 6**) using the same analysis as in **Table 3**.

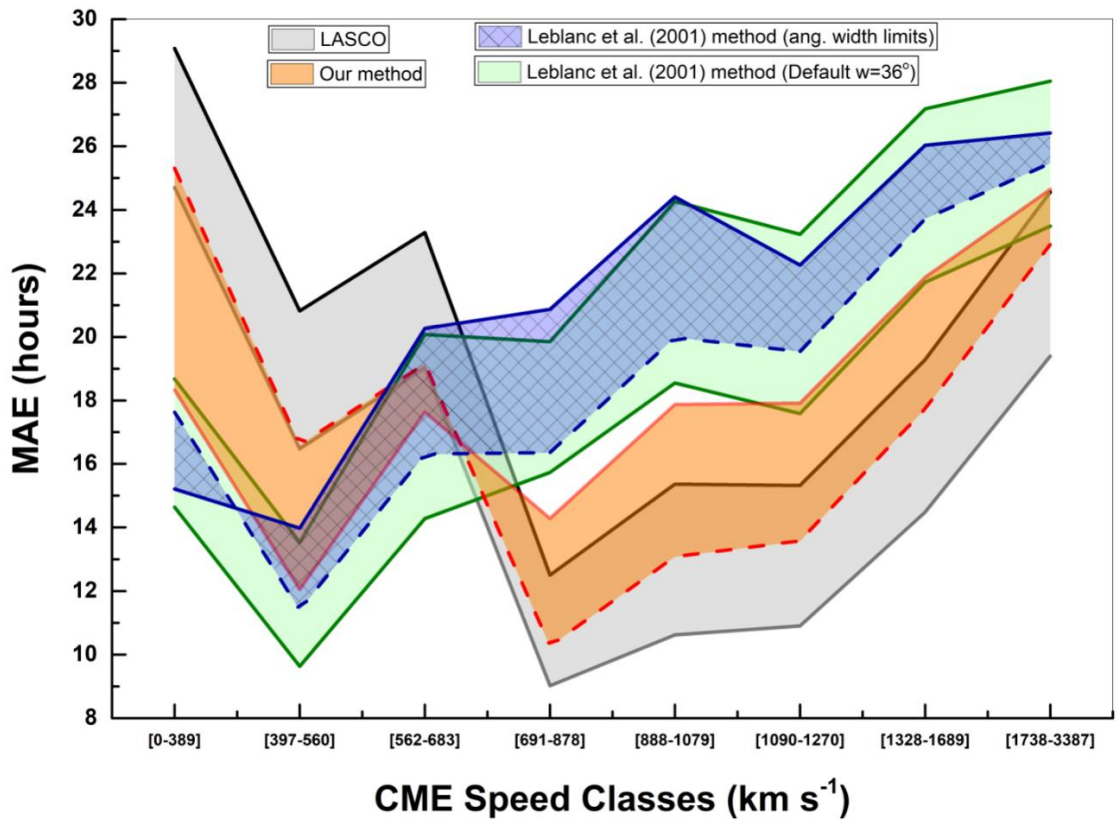

*Figure 6: Mean absolute error performance as a function of the CME LASCO POS speed for different speed correction approaches. The bands represent the MAE standard error. For the two methods using the upper/lower CME width to define the errors, the dashed (solid) line represent the lower (upper) width based results.*

**Figure 6** suggests that there exists some speed dependence on the ToA performance, as we surmised. In particular, the original Leblanc et al method outperforms all other methods for POS speeds in the [397, 560] km/s bin (MAE = 11.6 ± 2.0 hours) but it quickly deteriorates at higher speeds with the worst overall performance at the highest speed bin (>1730 km/s). Our method performs best for [691-878] km/s POS speeds (MAE=11.6 ±1.0 hours) and is generally on the low MAE end for higher speeds. Overall, Leblanc's method performs better for slower CMEs especially when $V_{POS}$ < 400 km/s, while our method performs better for faster CMEs with $V_{POS}$ > 700 km/s.

Additionally, our method performs better with the smaller CME angular width ($w_F = 41°$) for POS speeds < 683 km/s (solid lines in **Figure 6**) and with the larger CME width ($w_L = 94°$) for POS speeds > 683 km/s (dashed lines in **Figure 6**). This behavior suggests that fast CMEs are likely wider and hence are better represented by the shallow cone model and that slower CMEs are more closer to 3-part CMEs and hence are better represented by the ice-cream cones, as assumed by Leblanc. These results are consistent with higher expansion rates and the existence of a shock sheath surrounding the faster events as past analyses have shown (e.g. Vourlidas et al. 2013; 2017; Kwon et al. 2014; Balmaceda et al. 2018). It appears, therefore, that a hybrid approach for the speed deprojection, with an ice-cream or shallow cone models being used based on POS speed, may improve the ToA performance of empirical approaches (such as the EAMv2) particularly for medium to slow CMEs (i.e., <700 km/s). We also conclude that the IP propagation of these events is consistent with a smooth gradual deceleration profile.

On the other hand, **Figure 6** indicates, rather strongly that the POS speeds for fast events work just as well as the corrected speeds for ToA estimation. The only obvious improvement from our efforts is the reduced scatter in the MAE error. Why is that?

The easiest answer may be that the deprojection approach is simply inadequate. Relying CME-flare association and flare location for the corrections will be affected by CME deflections or source misidentification, particularly during high activity. Observations from off the Sun-Earth line and/or multi-viewpoint observations should, in principle, provide more reliable speeds. However, there seems to be a limit from that approach to the ToA accuracy, which stands at about 10 hours, as we mentioned before. In any case, since multi-viewpoint observations are uncertain in the future, we need to investigate how to improve the single-viewpoint, from the Sun-Earth line, observations that are forthcoming.

So, we see three possible obstacles for such observations. Firstly, the weak dependence of the ToA performance on the projection correction suggests that fast CMEs have a rather uniform speed across their front, which is rather flat.

Secondly, the linear speeds used as inputs (whether POS or corrected) are not sufficiently representative for the empirical characterization of the IP propagation. It is well known that fast CMEs decelerate strongly, particularly towards the edge of the LASCO FOV (e.g. Michalek et al. 2015, and references therein). Clearly, the use of a linear speed will introduce errors, and the errors will become more important with increasing CME speed. This behavior needs to be captured in empirical models but in lieu of that, we suggest to use the speed of a CME at 25 $R_{Sun}$ based on a $2^{nd}$ degree fit to the h-t points, for all events with linear POS speeds >700 km/s, at least.

A third possibility is that the method for estimating the ToA is not sensitive to the CME speed in the corona. The IP speed profile model is the other component in this methodology and assumes a smooth deceleration profile during the IP propagation of fast CMEs. This appears to an inadequate assumption but we do not pursue it further here. Suffice to say that our speculation is consistent with recent findings. For example, Wood et al (2015) find that a deceleration phase, followed by near-constant speeds coasting to 1 AU better described their events. Therefore, a two or three-phase description of the CME deceleration may lead to better ToA performance. To derive such an empirical model, however, it is necessary to obtain a sufficient number of CME kinematic measurements in the inner heliosphere over a representative time span of the solar cycle, covering both low and high activity phases. Such measurements should be available from the SECCHI/HI instruments and we plan to embark on such study next. Until such a model is constructed and tested, we do not expect any significant improvement in the ToA performance of empirical models.

## 4.  Summary and Conclusions

Correction of projection effects is very important for deriving reliable CME kinematic characteristics.

In this paper, we seek to improve the methodology for correcting the projection effects on speed measurements of CMEs from coronagraph observations. We focus on CMEs with known source locations thanks to previously identified flare associations. We start from widely used approaches that assume radial propagation from the flare site and a cone model for the CME. We introduce two variants to make the approach consistent with the current understanding of the CME structure. First, we assume that the CME front, as seen in the coronagraph images, is described by a shallow cone rather than a spherical cone (the so-called 'ice-cream' cone). This is a better description for energetic CMEs that exhibit a shock sheath around them. Secondly, as the CME width is an important but uncertain parameter (due to projection effects) in the projection correction, we bound that uncertainty by introducing an upper and lower limit to the actual CME width based on the statistical work of Balmaceda et al. (2018). In particular, we use the mean angular width for loop (L-Type) and flux rope (F-Type) CMEs (see Table 2 in Balmaceda et al., 2018) as the upper $w_L = w_{upp} = 94°$ and lower $w_F = w_{low} = 41°$ limit respectively. Thirdly, we correct for the location of the CME nose, consistently with the above assumptions. Our correction method is applied directly to the h-t measurements and the deprojected h-t points are then used to derive the deprojected CME speed with a linear fit. This approach provides naturally error bars for the derived speeds, which we can use in our ToA metrics. Our intention is to evaluate the effect of a particular speed correction as input to ToA empirical models in preparation for a future scenario of CME observations along the Sun-Earth line.

We apply this technique to 1037 CME-flare pairs extending across the last two solar cycles. Our deprojection technique results in reasonable speed ranges, a few km/s up to about 3500 km/s, in contrast to all previous methods, which tend to overcorrect CME speeds resulting in speeds in excess of 4000 km/s, which are energetically unfeasible. Our ratio of the deprojected to the POS CME speeds ranges from 1.0 to 1.37-2.86 for the angular width limits.

Since we use the individual h-t measurements, we can apply other selection criteria to enhance the robustness of the results, such as the number of h-t points. We conservatively assume that events with ten or more h-t points are more reliable for deriving speeds than events with fewer points. The corrected vs POS speeds for events with at least 10 h-t points show strong correlation (r=0.99) with a mean scatter of 15%. We plan to continue our analysis of the dataset and try to improve the speed estimates in a future publication. We test the ToA performance of our methods on a sample of 192 CME-ICME pairs (Paouris and Mavromichalaki, 2017b). We summarize our findings as follows:

- Our approach provides reasonable speeds across all source locations with maximum deprojected speeds of ~3500 km/s (**Figure 4**). This is clearly an improvement over all other deprojection methods, based on source region locations that tend to overcorrect speeds to unphysical levels and hence require special care for use in operational situations.

- We introduce an observationally-based method to bound the uncertainty in CME widths and speeds by assuming that the true CME width varies from 41° to 94° based on multi-viewpoint measurements.

- The analysis suggests that single viewpoint measurements are reliable (under our set of assumptions) since the mean difference between the corrected and the POS speeds for the full 1037 event sample is about 12.8%.

- Our speed correction methodology has superior performance to the 'ice cream' cone method of Leblanc et al. (2001) across all ToA metrics. We attribute this difference to the more reasonable corrected speeds and to the fact that CMEs, particularly faster events, are better described by a shallow cone (flatter front).

- We find only slight improvements in ToA metrics, with narrower dispersions (**Table 3**) when compared to using uncorrected speeds. This implies that the projected speeds are generally close to their true speeds (for the events considered here, at least), which in turn suggests a rather uniform speed profile across CME fronts in the middle corona. Therefore, the geometric correction plays a minor role in improving ToA prediction, at least when used with empirical models where the CME speed in the corona is the main input. In this view, coronagraphic h-t measurements may be insufficient for further improvement of the ToA, as discussed by Colaninno et al. (2013).

- Our analysis for halo CMEs ($w_p = 360°$), shows that our method results in MAE of 16.2±1.0 hours compared to 20.3±1.8 hours for the Leblanc et al method. This is a promising result since these CMEs often produce the most intense geomagnetic storms at Earth.

We gain further insight into the problem by examining the ToA performance as a function of CME POS speed. We split the 192-event sample evenly into 8 speed bins (24 events/bin).

- We find (**Figure 6**) a clear dependence of the ToA MAE with POS speed with the best performance resulting from events with slow to medium speeds (300-700 km/s).

- Our approach performs better for events > 700 km/s (MAE = 11.6) while the 'ice-cream' model performs better for events < 700 km/s (MAE = 11.6). Therefore, slower events are described better by a spherical front and faster events are likely flatter. This is consistent with the presence of a shock sheath for faster events.

- Given the modest improvement in ToA estimates from the deprojected speeds, we speculate that CME propagation may be a more important factor in driving the MAE uncertainty than the event geometry in the corona (see also Vourlidas et al. 2019). Our ToA estimates rely on an empirical kinematic model (EAMv2; Paouris & Mavromichalaki 2017b) that assumes a smooth deceleration profile during the CME IP propagation. The findings above indicate that the EAMv2 model performs well for events with POS speeds <700 km/s, so these events are likely experiencing a smoother deceleration than faster CMEs.

- The decreasing performance with increasing POS speed (above 700 km/s) suggests that the kinematic profile of fast CMEs is more complex than the assumption of gradual deceleration to 1 AU. A more nuanced empirical profile is required, possibly derived from heliospheric imaging observations off the Sun-Earth line.

- Another source of error is the linear fit to the LASCO h-t measurements since it is well known that fast CME decelerate rapidly. We suggest that the speed at $25R_{Sun}$, after the strongest deceleration has taken place, may be a better input to empirical models.

**Acknowledgements**

We are grateful to the providers of the solar data used in this work. The coronal mass ejection data are taken from the SOHO/LASCO CME list (http://cdaw.gsfc.nasa.gov/CME_list/). This CME catalog is generated and maintained at the CDAW Data Center by NASA and The Catholic University of America in cooperation with the Naval Research Laboratory. SOHO is a project of international cooperation between ESA and NASA. The soft X-ray data from GOES satellites are taken from the National Geophysical Data Center (ftp://ftp.ngdc.noaa.gov/STP/space-weather/solar-data/solar-features/solar-flares/x-rays/goes/xrs/).

E.P. acknowledges the State Scholarships Foundation of Greece. This research is co-financed by Greece and the European Union (European Social Fund - ESF) through the Operational Programme «Human Resources Development, Education and Lifelong Learning» in the context of the project "Reinforcement of Postdoctoral Researchers - 2nd Cycle" (MIS-5033021), implemented by the State Scholarships Foundation (IKY).

A.V. is supported by NASA grants NNX17AC47G and 80NSSC19K0069. A.P. and A.A. acknowledge the support through the ESA Contract No. 4000120480/NL/LF/hh" Solar Energetic Particle (SEP) Advanced Warning System (SAWS)". A.P. further acknowledges the TRACER project (http://members.noa.gr/atpapaio/tracer/), funded by the National Observatory of Athens (NOA) (Project ID: 5063).

**References:**

Aschwanden, M.J., Nitta, N.V., Wulser, J.-P., Lemen, J.R., Sandman, A., Vourlidas, A. & Colaninno, R.C., (2009). First measurements of the mass of coronal mass ejections from the EUV dimming observed with STEREO EUVI A+B spacecraft, *The Astrophysical Journal*, 706, 376–392. DOI:10.1088/0004-637X/706/1/376

Balmaceda, L.A., Vourlidas, A., Stenborg, G., & Dal Lago, A. (2018). How Reliable Are the Properties of Coronal Mass Ejections Measured from a Single Viewpoint?, *The Astrophysical Journal*, 863, 57. DOI:10.3847/1538-4357/aacff8

Billings, D.E. (1966). *A Guide to the Solar Corona*, Academic Press, New York.

Bronarska, K. & Michalek, G. (2018). Determination of projection effects of CMEs using quadrature observations with the two STEREO spacecraft, *Advances in Space Research*, 62, 2, 408-416. DOI: 10.1016/j.asr.2018.04.031

Brueckner, G.E., Howard, R.A., Koomen, M.J., Korendyke, C.M., Michels, D.J., Moses, J.D., Socker, D.G., Dere, K.P., Lamy, P.L., Llebaria, A., Bout, M.V., Schwenn, R., Simnett, G.M., Bedford D.K., & Eyles, C.J. (1995). The Large Angle Spectroscopic Coronagraph (LASCO), *Solar Physics*, 162, 357-402. DOI: 10.1007/BF00733434

Burkepile, J. T., Hundhausen, A. J., Stanger, A. L., St. Cyr, O. C., & Seiden, J.A. (2004). Role of projection effects on solar coronal mass ejection properties: 1. A study of CMEs associated with limb activity, *Journal of Geophysical Research*, 109, 3103. DOI: 10.1029/2003JA010149

Colaninno, R.C., Vourlidas, A., and Wu, C. C. (2013). Quantitative comparison of methods for predicting the arrival of coronal mass ejections at Earth based on multi view imaging, *Journal of Geophysical Research,* Space Physics, 118, 6866–6879, DOI: 10.1002/2013JA019205

Domingo, V., Fleck, B. & Poland, A.I. (1995). The SOHO mission: An overview, *Solar Physics, 162*, 1–37. DOI: 10.1007/BF00733425

Emslie, A. G., Dennis, B. R., Shih, A. Y., Chamberlin, P. C., Mewaldt, R. A., Moore, C. S., Share, G. H., Vourlidas, A., & Welsch, B. T. (2012). Global Energetics of Thirty-eight Large Solar Eruptive Events. *The Astrophysical Journal, 759*, 71. DOI: 10.1088/0004-637X/759/1/71

Gao, P. X., & Li, K. J., (2010). Velocity Distribution of CMEs After Projection Correction. *Chinese Astronomy and Astrophysics*, 34, 2, 154-162. DOI: 10.1016/j.chinastron.2010.04.003

Gonzalez, W.D., Dal Lago, A., Clua de Gonzalez, A.L., Vieira, L.E.A. & Tsurutani, B.T. (2004). Prediction of peak-Dst from halo CME/magnetic cloud-speed observations, *Journal of Atmospheric and Solar-Terrestrial Physics*, 66 (2), 161–165. DOI: 10.1016/j.jastp.2003.09.006

Gopalswamy, N., Yashiro, S., Michalek, G. Stenborg, G., Vourlidas, A., Freeland, S., & Howard, R. (2009). The SOHO/LASCO CME Catalog. *Earth, Moon, and Planets, 104*, 295–313. DOI: 10.1007/s11038-008-9282-7

Gopalswamy, N. (2016). History and development of coronal mass ejections as a key player in solar terrestrial relationship, *Geoscience Letters, 3*, 8. DOI: 10.1186/s40562-016-0039-2

Howard, T. A., Webb, D. F., Tappin, S. J., Mizuno, D. R., & Johnston, J. C. (2006). Tracking halo coronal mass ejections from 0 –1 AU and space weather forecasting using the Solar Mass Ejection Imager (SMEI), *Journal of Geophysical Research, 111*, A04105, doi: 10.1029/2005JA011349

Howard, T.A. & Tappin, S.J. (2008). Three-Dimensional Reconstruction of Two Solar Coronal Mass Ejections Using the STEREO Spacecraft, *Solar Physics, 252*, 373. DOI: 10.1007/s11207-008-9262-0

Howard, T. A., Nandy, D., & Koepke, A. C. (2008). Kinematic properties of solar coronal mass ejections: Correction for projection effects in spacecraft coronagraph measurements, *Journal of Geophysical Research*, *113*, A01104. doi:10.1029/2007JA012500

Jang, S., Moon, Y.J., Kim, R.S., Lee, H. & Cho, K.S. (2016). Comparison between 2D and 3D parameters of 306 front-side halo CMEs from 2009 to 2013, *The Astrophysical Journal*, 821:95. DOI: 10.3847/0004-637X/821/2/95

Kaiser, M.L., Kucera, T.A., Davila, J.M., St. Cyr, O. C., Guhathakurta, M. & Christian, E. (2008). The STEREO Mission: An Introduction. *Space Science Reviews, 136*, 5–16. DOI: 10.1007/s11214-007-9277-0

Kwon, R. Y., Zhang, J., Olmedo, O., (2014). New insights into the physical nature of coronal mass ejections and associated shock waves within the framework of the three-dimensional structure. *The Astrophysical Journal*, 794, 148, DOI: 10.1088/0004-637X/794/2/148

Kwon, R. Y., Zhang, J., & Vourlidas, A. (2015). Are halo-like solar coronal mass ejections merely a matter of geometric projection effects?, *The Astrophysical Journal Letters*, 799, 2. DOI: 10.1088/2041-8205/799/2/L29

Leblanc, Y., Dulk, G. A., Vourlidas, A., & Bougeret, J.-L. (2001). Tracing shock waves from the corona to 1 AU: Type II radio emission and relationship with CMEs, *Journal of Geophysical Research*, *106*, A11, 25301– 25312. DOI:10.1029/2000JA000260

Lee, H., Moon, Y.J., Na, H., Jang, S. & Lee, J.O. (2015). Are 3-D coronal mass ejection parameters from single-view observations consistent with multi view ones?, *Journal of Geophysical Research, Space Physics*, 120, 10, 237-10, 249. DOI: 10.1002/2015JA021118

Liewer, P., Panasenco, O., Vourlidas, A., & Colaninno, R. (2015). Observations and Analysis of the Non-Radial Propagation of Coronal Mass Ejections Near the Sun, *Solar Physics*, *290*, 3343. DOI: 10.1007/s11207-015-0794-9

Makela, P., Gopalswamy, N., & Yashiro, S. (2016) The radial speed-expansion speed relation for Earth-directed CMEs, *Space Weather, 14* (5), 368–378. DOI: 10.1002/2015SW001335

Mierla, M., Davila, J., Thompson, W., Inhester, B., Srivastava, N., Kramar, M., St. Cyr, O. C., Stenborg, G., & Howard, R. A. (2008). A Quick Method for Estimating the Propagation Direction of Coronal Mass Ejections Using STEREO-COR1 Images. *Solar Physics, 252*, 385–396. DOI: 10.1007/s11207-008-9267-8

Michalek, G., Gopalswamy, N., Yashiro, S., & Bronarska, K. (2015). Dynamics of CMEs in the LASCO Field of View. *Solar Physics, 290*, 903–917. DOI: 10.1007/s11207-015-0653-8

Michalek, G., Gopalswamy, N. & Yashiro, S. (2003). A new method for estimating widths, velocities, and source location of halo coronal mass ejections. *The Astrophysical Journal*, 584, 1. DOI: 10.1086/345526

Paouris, E., & Mavromichalaki, H. (2017a). Interplanetary Coronal Mass Ejections Resulting from Earth-Directed CMEs Using SOHO and ACE Combined Data During Solar Cycle 23, *Solar Physics, 292*, 30. DOI: 10.1007/s11207-017-1050-2

Paouris, E., & Mavromichalaki, H. (2017b), Effective Acceleration Model for the Arrival Time of Interplanetary Shocks driven by Coronal Mass Ejections, *Solar Physics, 292*, 180. DOI: 10.1007/s11207-017-1212-2

Papaioannou, A., Sandberg, I., Anastasiadis, A., Kouloumvakos, A., Georgoulis, M.K., Tziotziou, K., Tsiropoula, G., Jiggens, P., & Hilgers, A., (2016). Solar flares, coronal mass ejections and solar energetic particle event characteristics. *Journal of Space Weather and Space Climate, 6*, A42. DOI: 10.1051/swsc/2016035

Papaioannou, A., Anastasiadis, A., Sandberg, I., & Jiggens, P., (2018a). Nowcasting of Solar Energetic Particle Events using near real-time Coronal Mass Ejection characteristics in the framework of the FORSPEF tool. *Journal of Space Weather and Space Climate, 8*, A37. DOI: 10.1051/swsc/2018024

Papaioannou, A., Anastasiadis, A., Kouloumvakos, A., Paassilta, M., Vainio, R., Valtonen, E., Belov, A., Eroshenko, E., Abunina, M. & Abunin, A. (2018b). Nowcasting Solar Energetic Particle (SEP) Events using Principal Components Analysis (PCA). *Solar Physics, 293*, 100. DOI:10.1007/s11207-018-1320-7

Patsourakos, S., Vourlidas, A., Török, T., Kliem, B., Antiochos, S. K., Archontis, V., et al. (2020). Decoding the Pre-Eruptive Magnetic Field Configurations of Coronal Mass Ejections. Space Science Reviews, in print, *ArXiv:2010.10186 [Astro-Ph]*.

Schwenn, R. (2006). Space Weather: The Solar Perspective. *Living Reviews in Solar Physics, 3*, 2. DOI: 10.12942/lrsp-2006-2

Shen, C., Wang, Y., Pan, Z., Zhang, M., Ye, P., & Wang, S., (2013). Full halo coronal mass ejections: Do we need to correct the projection effect in terms of velocity?. *Journal of Geophysical Research Space Physics*, 118 (11), 6858–6865. DOI: 10.1002/2013JA018872

Temmer, M., Preiss, S. & Veronig, A.M. (2009). CME Projection Effects Studied with STEREO/COR and SOHO/LASCO. *Solar Physics, 256*, 183-199. DOI: 10.1007/s11207-009-9336-7

Thernisien, A., Vourlidas, A. & Howard, R. A. (2009). Forward Modeling of Coronal Mass Ejections Using STEREO/SECCHI Data. *Solar Physics, 256* (1-2), 111-130. DOI: 10.1007/s11207-009-9346-5

Verbeke, C., Mays, M. L., Temmer, M., Bingham, S., Steenburgh, R., Dumbovic, M., Nunez, M., Jian, L.K., Hess, P., Wiegand, C., Taktakishvili, A., & Andries, J., (2019). Benchmarking CME arrival time and impact: Progress on metadata, metrics, and events. *Space Weather, 17* (1), 6-26. DOI: 10.1029/2018SW002046

Vourlidas, A., Subramanian, P., Dere, K.P., & Howard, R.A., (2000). Large-Angle Spectrometric Coronagraph Measurements of the Energetics of Coronal Mass Ejections. *The Astrophysical Journal, 534*, 456-467. DOI: 10.1086/308747

Vourlidas, A., Lynch, B. J., Howard, R. A., & Li, Y. (2013). How Many CMEs Have Flux Ropes? Deciphering the Signatures of Shocks, Flux Ropes, and Prominences in Coronagraph Observations of CMEs. *Solar Physics, 284*, 179–201. DOI: 10.1007/s11207-012-0084-8

Vourlidas, A., Balmaceda, L. A., Stenborg, G., & Dal Lago, A. (2017). Multi-viewpoint Coronal Mass Ejection Catalog Based on STEREO COR2 Observations. *The Astrophysical Journal, 838* (2), 141. DOI: 10.3847/1538-4357/aa67f0

Vrsnak, B., Sudar, D., & Ruzdjak, D. (2005). The CME-flare relationship: are there really two types of CMEs?. *Astronomy and Astrophysics, 435* (3), 1149–1157. DOI: 10.1051/0004-6361:20042166

Vrsnak, B., Sudar, D., Ruzdjak, D., & Zic, T. (2007). Projection effects in coronal mass ejections. *Astronomy and Astrophysics, 469* (1), 339-346. DOI: 10.1051/0004-6361:20077175

Wood, B. E., Wu, C.-C., Lepping, R. P., Nieves-Chinchilla, T., Howard, R. A., Linton, M. G., & Socker, D. G. (2017). A STEREO survey of magnetic cloud coronal mass ejections observed at Earth in 2008–2012. *The Astrophysical Journal Supplement Series*, *229* (2), 29. DOI: 10.3847/1538-4365/229/2/29

Xie, H., Ofman, L., & Lawrence, G. (2004). Cone model for halo CMEs: Application to space weather forecasting. *Journal of Geophysical Research*, *109*, A03109. DOI: 10.1029/2003JA010226

Youssef, M. (2012). On the relation between the CMEs and the solar flares. *NRIAG Journal of Astronomy and Geophysics*, *1* (2), 172–178. DOI: 10.1016/j.nrjag.2012.12.014

Zhang, J., Dere, K.P., Howard, R.A., Kundu, M.R., & White, S.M. (2001). *The Astrophysical Journal, 559*, 452-462. DOI: 10.1086/322405

Zhang, J., Richardson, I.G., Webb, D.F., Gopalswamy, N., Huttunen, E., Kasper, J.C, Nitta, N.V., Poomvises, W., Thompson, B.J., Wu, C.C., Yashiro, S., & Zhukov, A.N. (2007). Solar and interplanetary sources of major geomagnetic storms (Dst < -100 nT) during 1996–2005. *Journal of Geophysical Research*, *112*, A10102. DOI: 10.1029/2007JA012321

Zhao, X. P., Plunkett, S. P., & Liu, W. (2002). Determination of geometrical and kinematical properties of halo coronal mass ejections using the cone model. *Journal of Geophysical Research*, 107 (A8). DOI: 10.1029/2001JA009143